\documentclass[prl,onecolumn,superscriptaddress]{revtex4}
\usepackage{amsmath,amssymb,mathrsfs}

\usepackage{hyperref}
\usepackage{paralist}
\usepackage{pdfpages}
\usepackage{lscape} %
 \usepackage{graphicx}

\def\be{\begin{equation}}
\def\ee{\end{equation}}
\def\bea{\begin{eqnarray}}
\def\eea{\end{eqnarray}}
\def\nn{\nonumber}

\newcommand{\Eq}[1]{Eq.~\eqref{#1}}

\newcommand{\beq}{\begin{equation}}
\newcommand{\eeq}{\end{equation}}

\newcommand{\beqa}{\begin{eqnarray}}
\newcommand{\eeqa}{\end{eqnarray}}

\newcommand{\Beqa}{\begin{eqnarray*}}
\newcommand{\Eeqa}{\end{eqnarray*}}

\DeclareMathOperator{\re}{Re}

\newcommand{\I}{\mathrm{i}}

\begin{document}

\title{Higher-order local and non-local correlations  for  1D strongly interacting Bose gas} 

\author{EJKP Nandani}
\affiliation{State Key Laboratory of Magnetic Resonance and Atomic and Molecular Physics,
Wuhan Institute of Physics and Mathematics, Chinese Academy of Sciences, Wuhan 430071, China.}
\affiliation{University of Chinese Academy of Sciences, Beijing 100049, China.}
\affiliation{Department of Mathematics,
University of Ruhuna,
Matara, 81000, Sri Lanka.}

\author{Rudolf A.\ R\"{o}mer}
\affiliation{Department of Physics and Center for Scientific Computing, University of Warwick, Coventry, CV4 7AL, UK.}

\author{Shina Tan}
\email[e-mail:]{shina.tan@physics.gatech.edu}
\affiliation{School of Physics, Georgia Institute of Technology, Atlanta, Georgia, 30332, USA}
\affiliation{Center for Cold Atom Physics, Chinese Academy of Sciences, Wuhan 430071, China.}

\author{Xi-Wen Guan}
\email[e-mail:]{xiwen.guan@anu.edu.au}
\affiliation{State Key Laboratory of Magnetic Resonance and Atomic and Molecular Physics,
Wuhan Institute of Physics and Mathematics, Chinese Academy of Sciences, Wuhan 430071, China.}
\affiliation{Department of Theoretical Physics, Research School of Physics and Engineering,
Australian National University, Canberra ACT 0200, Australia.}

\date{\today}

\begin{abstract}
 \paragraph*{ Abstract }
The correlation function is  an important quantity  in the physics of ultracold quantum gases because 
 it provides information about the quantum many-body wave function beyond the simple density
profile. In this paper we first study  the $M$-body local  correlation functions,   $g_M$, of the one-dimensional (1D)  strongly repulsive Bose gas within the Lieb-Liniger model using 
the  analytical method proposed  by Gangardt and Shlyapnikov \cite{dmg_gvs1,dmg_gvs}.
In the  strong repulsion regime  the 1D Bose gas at low temperatures is
equivalent to  a  gas of ideal particles obeying the non-mutual generalized exclusion statistics (GES) with a statistical parameter $\alpha =1-2/\gamma$,
i.e. the quasimomenta of $N$ strongly interacting bosons map to the momenta of $N$ free fermions via $k_i\approx \alpha k_i^F  $ with $i=1,\ldots, N$.  Here $\gamma$ is the dimensionless interaction strength within the Lieb-Liniger model. We rigorously prove that such a statistical parameter $\alpha$ solely determines the sub-leading order contribution
to the $M$-body local correlation function of the gas at strong but finite interaction strengths.
We explicitly calculate the correlation functions $g_M$  in terms of $\gamma$ and $\alpha$ at zero,  low, and intermediate temperatures.
For $M=2$ and $3$ our results reproduce the  known expressions for $g_{2}$ and $g_{3}$ with sub-leading terms
(see for instance  \cite{vadim,wang,Kormos:2009}).
We also express the leading order of the short distance \emph{non-local}  correlation functions 
$\langle\Psi^\dagger(x_1)\cdots\Psi^\dagger(x_M)\Psi(y_M)\cdots\Psi(y_1)\rangle$
of the strongly repulsive Bose gas
in terms of the wave function of $M$ bosons at zero collision energy and zero total momentum.
Here $\Psi(x)$ is the boson annihilation operator.
These  general formulas   of the higher-order local and non-local correlation functions of the 1D Bose gas  provide new insights into the many-body  physics.
 \paragraph*{keywords:}
 higher-order correlation functions, generalized exclusion statistics, Fermi distribution, Bethe Ansatz wave functions
\end{abstract}

\maketitle
%

\section{I. Introduction}

A fundamental principle of quantum statistical mechanics describes  two types of particles: bosons which satisfy the  Bose-Einstein statistics 
and fermions which   satisfy  the Fermi-Dirac statistics.  
An arbitrary number of identical bosons can occupy  one  quantum state whereas   no  more than one identical
fermion can occupy the same quantum state.
 The latter fundamental  concept  is called the ``Pauli Exclusion principle''.  
The statistics can be derived from the fact that the wave function of a system of bosons (fermions)  is symmetric (antisymmetric)  under the exchange of two particles. 
 However, under  certain conditions, a system of interacting bosons can be mapped into another system of fermions. A significant example is Girardeau's   Bose-Fermi mapping \cite{Girardeau:1960,Yukalov:2005} for the one-dimensional (1D) Lieb-Liniger Bose gas \cite{lieb} with an infinitely strong repulsion, which is called the Tonks-Girardeau gas \cite{Girardeau:1960,Tonks}. 
 This mapping was established based  on the observation that for an infinitely strong repulsion the relative wave function of the interacting  bosons must vanish when two bosons coincide spatially. Such behaviour mimics the Fermi statistics in identical fermions.  
 The Bose-Fermi mapping has tremendous applications in the study of strongly interacting quantum gases of ultracold atoms  \cite{Girardeau2,Zinner2, Santos, Pu, Levinsen, Zinner3, Levinsen2, Cui1}
 \footnote{The Bose-Fermi mapping may also be used in the reverse order. A prime example is the Usui transformation \cite{Usui}
 which maps fermion pairs to bosons.}.
 In this paper we present a new application of  Girardeau's  Bose-Fermi mapping to the study of higher-order local and non-local correlation functions. 
 
 An alternative  description of quantum statistics is  provided by Haldane's exclusion statistics~\cite{haldane,serguei}. Haldane formulated a description of 
 fractional statistics~\cite{haldane,serguei,yong} based on the 
 generalized Pauli exclusion principle, which counts the dimensions in the Hilbert space in a system with adding or removing an extra particle. 
It  is now called the generalized exclusion statistics (GES)~\cite{haldane}.
 In the strong coupling limit,  i.e.,  when the interaction strength  goes to infinity,   the Tonks-Girardeau gas is  in many ways equivalent to a noninteracting Fermi gas.
 In fact, the 1D $\delta$-function interacting bosons  can be mapped onto an ideal gas  \cite{yong} with the GES \cite{haldane} described by the statistical parameter $\alpha$.
 The equivalence between the 1D interacting bosons and the noninteracting particles
obeying the GES  is in general based on the  equivalence between the  thermodynamic Bethe Ansatz (TBA) equations~\cite{yang} and the GES equation \cite{yong,guan3}. The statistical profiles
and the thermodynamic properties
of the strongly interacting 1D Bose gas were studied  through the GES and TBA
approaches in Ref.~\cite{guan3}.
On the other hand, the statistical profiles of the strongly interacting 1D Bose gas at low temperatures are
equivalent to those of a gas of ideal particles obeying the non-mutual GES~\cite{guan3}, i.e. $\alpha$ is independent of the quasimomenta.
This equivalence has been recently investigated 
for a 1D model of interacting anyons \cite{guan3,guan2,anjan}. 
Such an equivalence between the 1D interacting Bose gas and the ideal gas with the GES  paves a way to calculating the correlation functions of the interacting system through the  ideal gas. In particular, in the non-mutual GES case,  we can map the quasimomenta of $N$ strongly interacting bosons  to the momenta of $N$ free fermions via $k_i\approx \alpha k_i^F  $ with $i=1,\ldots, N$, provided that the total momentum $k_1+\cdots+k_N=0$. Here $\alpha=1-2/\gamma$,
and $\gamma$ is the dimensionless interaction strength within the Lieb-Liniger model \cite{lieb}.

Correlation functions provide information about quantum many-body wave functions beyond the simple measurement of the density
profile~\cite{sykes}. Therefore, the study of $2$-body and $M$-body higher-order correlations is becoming an important theme in the physics of ultracold 
quantum gases~\cite{sykes}. The higher-order correlation was  first used by Hanbury Brown and Twiss (HBT) to measure the size of a distant binary star~\cite{yanli}.
Recently the non-local $M$-body correlations were measured with atomic  particles \cite{Dall:2013,Hodgman:2011}.
The local pair correlation function over a wide range of coupling strengths has been determined experimentally by measuring photoassociation
rates in the 1D Bose gas~\cite{toshiya,Kinoshita2004}.
Physically, the local pair correlation is a  measure of the probability of finding two particles at the same place.
Many studies have focused on  the local and non-local  correlations in 1D interacting uniform  Bose gases at zero and finite temperatures
 ~\cite{maxim,sykes,vladimir,holzmann,kv_dm_pd_gv,kozlowski,eckart,ovidiu,korepin,vadim,dmg_gvs,dmg_gvs1,oleksandr,Kormos:2010,Kormos:2011,Pozsgay:2011,Garcia:2014,Xu:2015,Guan:2011}. 
 Moreover, some groups have conducted the measurements of  the $2$-body and $3$-body correlations of bosons in 1D and 3D
 ~\cite{haller,tolra,toshiya,armijo}. Recently, people have studied the dynamics of strongly interacting bosons in 3D \cite{Makotyn:2014,Kira:2015}. 

In this paper,  we first  calculate the higher-order correlation functions of the 1D strongly interacting Bose gas by taking the asymptotic Bethe Ansatz wave function.
In light of an  analytical method developed by Gangardt and Shlyapnikov ~\cite{dmg_gvs,dmg_gvs1}, we  rigorously calculate the denominator and the numerator of the $M$-body correlation function up to the sub-leading order. Precisely speaking, the $M$-particle local correlations in the strong coupling limit ($\gamma\to\infty$) can be calculated through  the $M$-body correlation of free fermions by  using Wick's theorem and the Fermi-Dirac distribution.
However, for a strong but finite  interaction  the bosons do not exhibit pure Fermi statistics ~\cite{yong,guan3}.  
It is necessary  to consider a  correction to the pure Fermi statistics in the Gangardt/Shlyapnikov approach. 
It turns out that   for  strong but finite interaction  strengths the statistical parameter $\alpha$ solely determines the sub-leading order contribution to the $M$-body local correlation function.
In the strong coupling regime, the statistical profiles and the thermodynamic properties of the 1D Bose gas are equivalent to those of the ideal gas with the GES parameter $\alpha$~\cite{guan3}.
%
We derive explicit formulas of $g_M$ with sub-leading terms for  arbitrary $M=1,2\ldots$  at zero and  nonzero  temperatures.
For the special cases of $M=2,\,3$, our results reduce to the known results of $g_2$ and $g_3$ with the sub-leading terms as given in
Refs.~\cite{dmg_gvs,dmg_gvs1,Kheruntsyan,Cazalilla,Kormos:2009,Kormos:2010}.
Furthermore,  we analytically calculate
the leading order of the short distance  $M$-body non-local  correlation functions 
$\langle\Psi^\dagger(x_1)\cdots\Psi^\dagger(x_M)\Psi(y_M)\cdots\Psi(y_1)\rangle$ of
the  1D  strongly repulsive  Bose gas. Here $\Psi(x)$ is  the boson annihilation operator.

Our paper is  organized as follows. 
In  Section II we  derive a general formula for  the $M$-body local correlation function $g_M$
of 1D bosons at a large interaction strength, $\gamma\gg1$. In Section III, we analytically calculate the $M$-body local correlation at various temperatures.
We then compare our results for $g_M$ with the previous results for $g_2$ and $g_3$  by  Gangardt and Shlyapnikov \cite{dmg_gvs},
Vadim {\em et.al.} \cite{vadim}, Kormos \textit{et.al.} \cite{Kormos:2009}, and Wang {\em et.al.}~\cite{wang}.
  In Section IV we study the wave function of $M$ interacting bosons at zero collision energy. 
  In Section V we calculate the short distance  $M$-body non-local  correlation functions of the ideal Fermi gas. 
   In Section VI we determine   the short distance  $M$-body non-local  correlation functions of the 1D strongly repulsive Bose gas,
   expressing such correlation functions in terms of the wave functions defined in Section IV.
In Section VII we conclude.  


\section{II. The higher-order local correlation functions of 1D Bosons}
\label{sec2}

We consider $N$ bosons  interacting via repulsive $\delta$-function potentials in 1D with Hamiltonian
\begin{equation}\label{hamiltonian}
 H=\dfrac{\hbar^{2}}{2m}\left(\sum\limits_{j=1}^{N}-\partial^{2}_{x_{j}}+ 2c\sum\limits_{j>l}\delta(x_{j}-x_{l})\right),
\end{equation}
where $m$ is the mass of each boson, $x_{j}$ is the coordinate of the $j$th boson and $c>0$ is the coupling constant~\cite{lieb}.
The Hamiltonian \eqref{hamiltonian} is diagonalized by means of the Bethe Ansatz~\cite{lieb,dmg_gvs,guan5}. For convenience we define the dimensionless interaction strength $\gamma =c/n$, where $n=N/L$ is the number density of the bosons. 
Assuming the periodic boundary condition, $\psi(0,x_{2},\ldots,x_{N})=\psi(x_{2},\ldots,x_N,L)$,
we have  the energy eigenfunction \cite{lieb,dmg_gvs,gaudin1}
\begin{align}
 \psi(x_{1},x_{2},\ldots,x_{N})
 &=\Big[\prod_{1\le i<j\le N}\Big(1+\frac{\partial_{x_j}-\partial_{x_i}}{c}\Big)\Big]\phi^{(0)}(x_1,x_2,\ldots,x_N)\nn\\
 &=\sum_p(-1)^p\Big[\prod_{1\le i<j\le N}\Big(1+\frac{\I k_{pj}-\I k_{pi}}{c}\Big)\Big]\exp\Big(\sum_{j=1}^N\I k_{pj}x_j\Big)\label{wave}
\end{align}
in the domain  $0\le x_{1}\le x_{2}\le\ldots\le x_{N}\le L$, where
\beq\label{phi0}
\phi^{(0)}(x_1,x_2,\ldots,x_N)\equiv\sum_p(-1)^p\exp\Big(\sum_{j=1}^N\I k_{pj}x_j\Big),\text{ for all $x$'s}
\eeq
is a completely antisymmetric function, and $k_1,\ldots,k_N$ are the quasimomenta \cite{lieb}.
Without loss of generality we shall assume that $k_1<k_2<\cdots<k_N$.
The sums in \Eq{wave} and \Eq{phi0} run over $N!$ permutations of the integers $1,\ldots,N$,
and $(-1)^p =+ 1$ ($-1$) for an even (odd) permutation.
%
 The $M$-particle local correlation function is defined as~\cite{dmg_gvs}
 \begin{eqnarray}
   g_{M}&=&\dfrac{\langle\psi|(\Psi^{\dagger}(0))^{M}(\Psi(0))^{M}|\psi\rangle}{\langle\psi|\psi\rangle}=\langle(\Psi^{\dagger}(0))^{M}(\Psi(0))^{M}\rangle\nn\\
   &=&\frac{N!}{(N-M)!}\frac{\int | \psi\left(0,\cdots,0, x_{M+1},\cdots, x_N \right)|^2dx_{M+1}\cdots dx_N}{\int | \psi\left(x_{1},\cdots, x_N \right)|^2dx_{1}\cdots dx_N},\label{corre-f}
 \end{eqnarray}
where $|\psi\rangle$ is the $N$-body energy eigenstate associated with the wave function $\psi$, and
$\Psi^{\dagger}(x)$ and $\Psi(x)$ are respectively the creation and the annihilation operators of the bosons.
The evaluations of the numerator and the denominator in \Eq{corre-f} are extremely hard even for  the strong coupling regime. 
In order to work  out $g_M$ we need to expand both the numerator and the denominator to the sub-leading order in the large coupling limit. 
After lengthy calculations, detailed in Appendix I, we find
\begin{align}
\int_0^L \Big| \psi\left(x_{1},\cdots, x_N \right)\Big|^2dx_{1}\cdots dx_N&=\left[ 1+\frac{2N(N-1)}{cL}+O(c^{-2}) \right]\int_0^L|\phi^{(0)}(x_1,\cdots, x_N)|^2dx_1\cdots dx_N\nn\\
&=\big[\alpha^{1-N}+O(c^{-2})\big]\int_0^L|\phi^{(0)}(x_1,\cdots, x_N)|^2dx_1\cdots dx_N,\label{denominator}
\end{align}
\begin{align}
& \int_0^L | \psi\left(0,\cdots,0, x_{M+1},\cdots, x_N \right)|^2dx_{M+1}\cdots dx_N\nn \\
&=c^{-M(M-1)} \int_0^L\Big| \phi^{(\Delta)}\left(0, x_{M+1},\cdots, x_N \right)\Big|^2dx_{M+1}\cdots dx_N+O(c^{-M(M-1)-2}),\label{numerator}
\end{align}
where
\beq
 \phi^{(\Delta)}(0,x_{M+1},\ldots,x_{N}) \equiv\Big[\prod_{1\le i<j\le M}(\partial_{x_j}-\partial_{x_i})\Big]\phi^{(0)}(x_{1},\ldots,x_{N})\Big|_{x_{1}=\cdots=x_{M}=0}.
\eeq
The quasimomenta $k_1, k_2, \ldots, k_N$ deviate from pure Fermi statistics at a large but finite interaction strength.
Instead, they obey the non-mutual GES \cite{yong,guan3}. The deviation from Fermi statistics for a large but finite
interaction strength $\gamma$ is described by the non-mutual  GES  parameter  $\alpha=1-2/\gamma$ ~\cite{guan3}.
 If the total momentum $k_1+\cdots+k_N=0$, we have $k_i=k_i^F\alpha+O(c^{-2})$, where
 $k_i^F=2\pi m_i/L$ and the $m_i$'s are integers satisfying $m_1<m_2<\cdots<m_N$.

In the strong coupling limit, $\gamma \gg1$, making a scaling change  $x_i= x_i^F/\alpha $ with $i=1,\ldots, N$, we rewrite the numerator in \Eq{corre-f} as
\begin{align}
&\int_0^L \Big| \psi\left(0,\cdots,0, x_{M+1},\cdots, x_N \right)\Big|^2dx_{M+1}\cdots dx_N\nn\\
&=c^{-M(M-1)}\int_0^L\bigg|\Big[\prod_{1\le i<j\le M}(\partial_{x_j}-\partial_{x_i})\Big]\sum_p(-1)^p\exp\Big(\sum_{j=1}^N\I k_{pj}x_j\Big)\Big|_{x_1=\cdots=x_M=0}\bigg|^2
dx_{M+1}\cdots dx_N+O(c^{-M(M-1)-2})\nn\\
&=c^{-M(M-1)}\int_0^{\alpha L}\bigg|\Big[\prod_{1\le i<j\le M}\alpha(\partial_{x_j^F}-\partial_{x_i^F})\Big]\sum_p(-1)^p\exp\Big(\sum_{j=1}^N\I k_{pj}^Fx_j^F\Big)\Big|_{x_1^F=\cdots=x_M^F=0}\bigg|^2
\alpha^{M-N}dx_{M+1}^F\cdots dx_N^F\nn\\
&\quad+O(c^{-M(M-1)-2})\nn\\
&=\frac{\alpha^{M^2-N}}{c^{M(M-1)}}\int_{0}^{\alpha L}\Big|\phi^{(\Delta F)}(0,x_{M+1}^F,\cdots,x_N^F)\Big|^2dx_{M+1}^F\cdots dx_N^F+O(c^{-M(M-1)-2}),
\end{align}
where $\partial_{x_i^F}$ is the partial derivative with respect to $x_i^F$,
\beq
\phi^{(\Delta F)}(0,x_{M+1}^F,\cdots,x_N^F)\equiv\Big[\prod_{1\le i<j\le M}(\partial_{x_j^F}-\partial_{x_i^F})\Big]\phi^F(x_1^F,\cdots,x_N^F)\Big|_{x_1^F=\cdots=x_M^F=0},
\eeq
and $\phi^F$ is the wave function of ideal fermions:
\beq
\phi^F(x_1^F,\cdots,x_N^F)\equiv\sum_p(-1)^p\exp\Big(\sum_{j=1}^N\I k_{pj}^Fx_j^F\Big).
\eeq
It is easy to see that whenever any one of the $(N-M)$ arguments $x_{M+1}^F,\ldots,x_N^F$ goes to zero,
say $x_i^F\to0$, for some $i$ that satisfies $M+1\le i\le N$, the function $\phi^{(\Delta F)}(0,x_{M+1}^F,\cdots,x_N^F)$ goes to zero like $(x_i^F)^M$.
On the other hand, the function $\phi^{(\Delta F)}(0,x_{M+1}^F,\cdots,x_N^F)$ is periodic:
\beq
\phi^{(\Delta F)}(0,x_{M+1}^F,\cdots,x_{i-1}^F,x_i^F+L,x_{i+1}^F,\cdots,x_N^F)=\phi^{(\Delta F)}(0,x_{M+1}^F,\cdots,x_{i-1}^F,x_i^F,x_{i+1}^F,\cdots,x_N^F).
\eeq
Thus, whenever $\alpha L<x_i^F<L$ for some $i$ satisfying $M+1\le i\le N$, the function
$\phi^{(\Delta F)}(0,x_{M+1}^F,\cdots,x_N^F)$ is of the order $c^{-M}$. So
\beq
\int_{0}^{\alpha L}\Big|\phi^{(\Delta F)}(0,x_{M+1}^F,\cdots,x_N^F)\Big|^2dx_{M+1}^F\cdots dx_N^F
=\int_{0}^{L}\Big|\phi^{(\Delta F)}(0,x_{M+1}^F,\cdots,x_N^F)\Big|^2dx_{M+1}^F\cdots dx_N^F+O(c^{-2M}).
\eeq
Assuming that $M\ge 1$, we thus find
\begin{align}
\int_0^L \Big| \psi\left(0,\cdots,0, x_{M+1},\cdots, x_N \right)\Big|^2dx_{M+1}\cdots dx_N
&=\frac{\alpha^{M^2-N}}{c^{M(M-1)}}\int_{0}^{L}\Big|\phi^{(\Delta F)}(0,x_{M+1}^F,\cdots,x_N^F)\Big|^2dx_{M+1}^F\cdots dx_N^F\nn\\
&\quad+O(c^{-M(M-1)-2}).\label{numerator2}
\end{align}
A brute-force calculation yields
\beq
\int_0^L|\phi^{(0)}(x_1,\cdots,x_N)|^2dx_1\cdots dx_N=N!\,L^N\det S,
\eeq
where $S$ is and $N\times N$ matrix with elements
\beq
S_{ij}=\mathrm{sinc}\frac{(k_i-k_j)L}{2},
\eeq
where
\begin{equation*}
\mathrm{sinc}(\xi)\equiv \left\{ \begin{array}{ll} \frac{{\rm sin} \xi }{\xi},&\xi\ne 0,\\
1, & \xi=0.
\end{array}\right.
\end{equation*}
Since $S_{ij}=1$ for $i=j$, and $S_{ij}=O(1/c)$ for $i\ne j$, we find
\beq
\det S=1+O(c^{-2}).
\eeq
Thus
\beq
\int_0^L|\phi^{(0)}(x_1,\cdots,x_N)|^2dx_1\cdots dx_N=\big[1+O(c^{-2})\big]N!\,L^N
=\big[1+O(c^{-2})\big]\int_0^L|\phi^{F}(x_1^F,\cdots,x_N^F)|^2dx_1^F\cdots dx_N^F.
\eeq
Substituting the above formula into \Eq{denominator}, we get
\beq\label{denominator2}
\int_0^L|\psi(x_1,\cdots,x_N)|^2dx_1\cdots dx_N=\big[\alpha^{1-N}+O(c^{-2})\big]\int_0^L|\phi^F(x_1^F,\cdots,x_N^F)|^2dx_1^F\cdots dx_N^F.
\eeq
Substituting  \Eq{numerator2} and \Eq{denominator2} into \Eq{corre-f}, we find
\beq
g_M=\frac{\alpha^{M^2-1}}{c^{M(M-1)}}\frac{N!}{(N-M)!}\frac{\int_0^L|\phi^{(\Delta F)}(0,x_{M+1},\cdots,x_N)|^2dx_{M+1}\cdots dx_N}
{\int_0^L|\phi^F(x_1,\cdots,x_N)|^2dx_1\cdots dx_N}+O(c^{-M(M-1)-2}).\label{corre-f2}
\eeq
From the definition of $\phi^{(\Delta F)}$, we find
\begin{align}
&\frac{N!}{(N-M)!}\frac{\int_0^L|\phi^{(\Delta F)}(0,x_{M+1},\cdots,x_N)|^2dx_{M+1}\cdots dx_N}
{\int_0^L|\phi^F(x_1,\cdots,x_N)|^2dx_1\cdots dx_N}\nn\\
&=\Delta_M(\partial_x)\Delta_M(\partial_y)
\frac{N!\int_0^L\phi^{F*}(x_1,\cdots,x_N)\phi^F(y_1,\cdots,y_M,x_{M+1},\cdots,x_N)dx_{M+1}\cdots dx_N}
{(N-M)!\int_0^L|\phi^F(z_1,\cdots,z_N)|^2dz_1\cdots dz_N}\Big|_{x_1=\cdots=x_M=y_1=\cdots=y_M=0},\label{DeltaFsquareintegral}
\end{align}
where $\Delta_M(\partial_x)\equiv\Delta_M(\partial_{x_1},\cdots,\partial_{x_M})$, and
\beq
\Delta_M(\xi_1,\cdots,\xi_M)\equiv\prod_{1\le i<j\le M}(\xi_j-\xi_i)=\det \Xi
\eeq
is the Vandermonde determinant. Here $\Xi$ is an $M\times M$ matrix with matrix elements $M_{ij}=\xi_i^{j-1}$ ($1\le i,j\le M$).

Since $\phi^F(x_1,\cdots,x_N)$ is a Slater determinant, it satisfies Wick's theorem:
\beq\label{Wick}
\frac{N!\int_0^L\phi^{F*}(x_1,\cdots,x_N)\phi^F(y_1,\cdots,y_M,x_{M+1},\cdots,x_N)dx_{M+1}\cdots dx_N}
{(N-M)!\int_0^L|\phi^F(z_1,\cdots,z_N)|^2dz_1\cdots dz_N}
=\sum_q(-1)^qG(x_1,y_{q1})G(x_2,y_{q2})\cdots G(x_M,y_{qM}),
\eeq
where the sum runs over all the $M!$ permutations of the integers $1,\cdots,M$, and
\beq
G(x,y)\equiv\frac{N\int_0^L\phi^{F^*}(x,x_2,\cdots,x_N)\phi^F(y,x_2,\cdots,x_N)dx_2\cdots dx_N}{\int_0^L|\phi^F(z_1,\cdots,z_N)|^2dz_1\cdots dz_N}
\eeq
is the 1-particle reduced density matrix of ideal fermions. Substituting the definition of $\phi^F$, we find
\beq
G(x,y)=\frac{1}{L}\sum_{i=1}^N\exp\big[-\I k_i^F(x-y)\big].
\eeq
Substituting the above formula into \Eq{Wick}, and then \Eq{Wick} into \Eq{DeltaFsquareintegral}, we find
\beq
\frac{N!}{(N-M)!}\frac{\int_0^L|\phi^{(\Delta F)}(0,x_{M+1},\cdots,x_N)|^2dx_{M+1}\cdots dx_N}
{\int_0^L|\phi^F(x_1,\cdots,x_N)|^2dx_1\cdots dx_N}
=\frac{M!}{L^M}\sum_{i_1=1}^N\cdots\sum_{i_M=1}^N\Delta_M^2(k_{i_1}^F,\cdots,k_{i_M}^F).
\label{DeltaFsquareintegral2}
\eeq
Substituting this result into \Eq{corre-f2}, we find
\beq\label{corre-f3}
g_M=\frac{\alpha^{M^2-1}M!}{c^{M(M-1)}L^M}\sum_{i_1=1}^N\cdots\sum_{i_M=1}^N\Delta_M^2(k_{i_1}^F,\cdots,k_{i_M}^F)+O(c^{-M(M-1)-2}).
\eeq
In the thermodynamic limit, the fermionic momenta $k_1^F,\cdots,k_N^F$ obey the Fermi distribution:
in the interval of momenta $(k,k+dk)$, where $dk$ is large compared to $2\pi/L$ but small compared to $n$,
the number of fermionic momenta is $\frac{Ldk}{2\pi}f(k)$, where
\beq
f(k)=\frac{1}{1+e^{\frac{\hbar^2k^2/2m-\mu}{k_BT}}},
\eeq
$k_B$ is Boltzmann's constant,
and $T$ and $\mu$ are respectively the temperature and the chemical potential of the 1D Bose gas \emph{after it is tuned to infinite coupling adiabatically}
at a fixed density $n$. At strong coupling, the actual temperature is very close to $T$.
In the thermodynamic limit, we thus find
\beq\label{corre-f4}
g_M=\frac{M!\,\alpha^{M^2-1}}{(2\pi)^Mc^{M(M-1)}}\int_{-\infty}^\infty dp_1\cdots dp_M\,f(p_1)\cdots f(p_M)\Delta_M^2(p_1,\cdots,p_M)+O(c^{-M(M-1)-2}).
\eeq
We would like to mention that the integral form of \Eq{corre-f4} with $\alpha$ set to 1 was derived in Refs.~\cite{dmg_gvs,Pozsgay:2011}. Here the quantum statistical correction $\alpha=1-2/\gamma$ contributes to the subleading term of the high order correlation function; see the proof given in the Appendix. 
At zero temperature $f(k)=\Theta(k_F-|k|)$, where $\Theta(\xi)$ is the Heaviside step function
and the Fermi-like momentum $k_F=\pi n$ \cite{guan3}. At nonzero temperatures $f(k)$ is broadened. Note also that
\beq\label{f(k)normalization}
N=\int_{-\infty}^\infty\frac{Ldk}{2\pi}f(k).
\eeq
Making a change of variable
\beq\label{kx}
k=2\pi nz,
\eeq
we obtain
\beq\label{corre9}
\frac{g_M}{n^M}=M!\Big(\frac{2\pi}{\gamma}\Big)^{M(M-1)}\alpha^{M^2-1}\int_{-\infty}^\infty dz_1\cdots dz_M\,N(z_1)\cdots N(z_M)\Delta_M^2(z_1,\cdots,z_M)
+O(\gamma^{-M(M-1)-2}),
\eeq
where
\beq
N(z)\equiv f(2\pi nz).
\eeq
Equation~\eqref{corre9} will be  used later to calculate  the higher-order local correlation functions.

\section{III. The Higher-order local correlations at various temperatures}

\subsection{General considerations}
From \Eq{f(k)normalization}, we obtain 
\begin{equation}\label{norm}
 \int_{-\infty}^{\infty}N(z)dz=1.
 \end{equation}
The multiple integral on the right hand side of Eq.\ \eqref{corre9} can be calculated by using the orthogonal polynomials in the random matrix model~\cite{mehta,freilikher},
yielding  
 \begin{equation}\label{corre8}
 \int_{-\infty}^{\infty}dz_{1}\cdots dz_{M}\,N(z_1)\cdots N(z_M)\Delta_{M}^{2}(z_1,\cdots,z_M) = M!\prod_{j=0}^{M-1}h_{j},
 \end{equation}
 where $h_{j}$ ($j=0,1,2,\ldots$) are the norm-squares of the monic orthogonal polynomials $P_{i}(z)$ with weight function $N(z)$ 
 \begin{equation}\label{hj}
 \int_{-\infty}^{\infty}P_{i}(z)P_{j}(z)N(z)dz =h_{j}\delta_{ij}.
\end{equation}
The  monic orthogonal polynomials can be found by using the Gram-Schmidt process 
 \begin{equation}
  P_{j}(z)=z^{j}-\sum\limits_{i=0}^{j-1}\dfrac{\langle z^{j},P_{i}(z)\rangle}{\langle P_{i}(z),P_{i}(z)\rangle}P_{i}(z), 
 \end{equation}
where $\langle A(z),B(z)\rangle =\int_{-a}^b A(z)B(z)N(z) dz$ within a finite range $-\infty<a<b<\infty$.
 Therefore, \Eq{corre9} can be re-expressed as
\begin{equation}
\frac{g_M}{n^M}=(M!)^{2}\left(\dfrac{2\pi}{\gamma}\right)^{M(M-1)}\alpha^{M^{2}-1}\prod_{j=0}^{M-1}h_{j}+O(\gamma^{-M(M-1)-2}).\label{g-M-G}
\end{equation}
We shall use \Eq{g-M-G} to calculate the $M$-particle local correlation function at various temperatures.

We define
\beq
D_{ij}\equiv\int_{-\infty}^\infty z^{i+j}N(z)dz\equiv D_{i+j},
\eeq
where $i$ and $j$ are nonnegative integers.
Because $N(z)$ is an even function, $D_{ij}=0$ if $i+j$ is odd.
We may expand the monic orthogonal polynomials as
\beq
P_i(z)=\sum_{j=0}^iP_{ij}z^j,
\eeq
where $P_{ii}\equiv 1$. Substituting this formula into \Eq{hj}, we find
\beq
\sum_{k=0}^i\sum_{l=0}^jP_{ik}D_{kl}P_{jl}=h_i\delta_{ij},
\eeq
which may be written in the matrix form:
\beq\label{PDP}
PDP^T=h,
\eeq
where $h$ is a diagonal matrix with diagonal elements $h_0,h_1,\cdots$, and $P$ is a lower triangular matrix whose diagonal elements are all 1.
Note that in the above equation, the row number and the column number of each matrix starts from 0.
Given the matrix $D$, we can solve \Eq{PDP} to find $P_{ij}$ and $h_i$.

\subsection{Zero temperature}
At zero temperature, strongly interacting 1D bosons have a Fermi-like surface \cite{guan3} 
\begin{equation}\label{zero}
 N(z)=\Theta\Big(\frac{1}{2}-|z|\Big).
      \end{equation}
Thus \Eq{hj} is simplified as
 \begin{equation}
 \int_{-1/2}^{1/2}P_{i}(z)P_{j}(z)dz =h_{j}\delta_{ij}.\label{hjT=0}
\end{equation}
We can express $P_j(z)$ in terms of the Legendre polynomials
 \begin{equation}
  Q_j(x)\equiv\sum\limits_{k=0}^{[j/2]}\dfrac{(2j-2k)!(-1)^{k}}{2^{j}(j-k)!k!(j-2k)!}x^{j-2k},
 \end{equation}
 which satisfy the orthogonality condition in the interval $(-1,+1)$:
 \beq
 \int_{-1}^{+1}Q_i(x)Q_j(x)dx=\frac{2}{2j+1}\delta_{ij}.
 \eeq
Comparing the properties of $P_j(z)$ and those of $Q_j(x)$, we find
\beq
P_j(z)=\frac{(j!)^2}{(2j)!}Q_j(2z)
\eeq
at zero temperature. Therefore
\beq
h_j=\int_{-1/2}^{1/2}\big[P_j(z)\big]^2dz=\frac{1}{2j+1}\bigg[\frac{(j!)^2}{(2j)!}\bigg]^2.
\eeq
Substituting the above result into \Eq{g-M-G}, we find the $M$-particle local correlation function at zero temperature 
\begin{eqnarray}
g_{M}
&=&n^{M}\left(\dfrac{\pi}{\gamma}\right)^{M(M-1)}\dfrac{\alpha^{(M^{2}-1)}}{(2M-1)!!}\left[\dfrac{\prod_{j=1}^{M}(j!)}{\prod_{j=1}^{M-1}(2j-1)!!}\right]^{2}
+O(\gamma^{-M(M-1)-2}). \label{M-proof}
 \end{eqnarray}
 The above formula is accurate at the leading and sub-leading orders in $1/\gamma$.

\subsection{Low temperatures, $T\ll T_d$}
The Sommerfeld expansion is applied to the evaluation of the norm-squares of the monic orthogonal polynomials at low temperatures.
The general expression for the moments of the distribution can be expressed as
\beq\label{DlowT}
D_{2j}=\frac{1}{ 2^{2j+1}}\left[\dfrac{2}{2j+1}+\dfrac{j}{6}\left(\dfrac{\tau}{\pi}\right)^{2}+o(\tau^2)\right],
 \text{ if }T\ll T_d,
\eeq
where $j$ is any nonnegative integer, $\tau\equiv T/T_d$, and $T_{d}\equiv\hbar^{2}n^{2}/(2mk_B)$ is the quantum degeneracy temperature.

At low temperatures $T\ll T_d$, solving \Eq{PDP} using \Eq{DlowT}, we find
\beq
h_i=\frac{1}{2i+1}\Big[\frac{(i!)^2}{(2i)!}\Big]^2\Big[1+\frac{i(i+1)(2i+1)\tau^2}{24\pi^2}+o(\tau^2)\Big].
\eeq
Substituting this result into \Eq{g-M-G}, we find
 \beq
\frac{g^M}{n^M}=\frac{\big(\prod_{j=1}^Mj!\big)^2\alpha^{(M^{2}-1)}}{\big[\prod_{j=1}^{M-1}(2j-1)!!\big]^2(2M-1)!!}\Big(\frac{\pi}{\gamma}\Big)^{M(M-1)}\left[ 1+ \frac{1}{48} M^2 (M^2-1)\left(\dfrac{\tau}{\pi}\right)^{2}+o(\tau^2)\right] 
+O(\gamma^{-M(M-1)-2})\label{correlation-C} .
\eeq
Equation~\eqref{correlation-C} reduces to \Eq{M-proof} at zero temperature.
We would like to emphasize that the above explicit expression of the $M$-body correlation functions contains the leading and sub-leading order terms in $1/\gamma$. 
It is very interesting to observe the many-body correlation effects with respect to the interaction parameter $\gamma$, statistics parameter $ \alpha$,
and the reduced temperature $\tau$. We also see that the thermal fluctuation strongly affects the higher-order correlation functions.

\subsection{High temperatures, $T_d\ll T\ll\gamma^2T_d$}

In Ref.~\cite{dmg_gvs} it was noticed that at temperatures $T\gg T_{d}$, the characteristic momentum of the particles is the thermal momentum $K_T\sim1/\Lambda$,
where $\Lambda=(2\pi\hbar^{2}/mk_{B}T)^{1/2}$ is the thermal de Broglie wavelength. Therefore, the small 
parameter for the expansion of the amplitudes in Eq.~\eqref{wave} is $1/\Lambda c$. Thus, it  must satisfy the inequality $1/\Lambda c\ll1$, which requires $T\ll\gamma^{2}T_{d}$.
At high temperatures $T\gg T_d$, the thermal wave length $\Lambda$ is much smaller than the average distance between two particles and  the system approaches a Maxwell-Boltzman distribution~\cite{guan3,guan5}.
In the Boltzmann limit $T\gg T_d$, we have $\zeta^{'}= e^{(\varepsilon(z)-\mu)/k_{B}T}\gg1$, where $\varepsilon(z)\equiv\hbar^2(2\pi nz)^2/2m$, so
we may simply approximate $N(z)$ as~\cite{wung,yong}
 $N(z)\simeq \dfrac{1}{\zeta^{'}}= \dfrac{1}{e^{(\varepsilon(z)-\mu)/k_{B}T}}.$
In the temperature regime $T_d\ll T\ll \gamma^2T_d$, we thus have
\beq\label{DhighT}
D_{2j}=\frac{(2j-1)!!}{8^j\pi^{2j}}\tau^{j}.
\eeq
Solving \Eq{PDP} using \Eq{DhighT}, we find
\beq
h_i=\frac{i!\,\tau^i}{2^{3i}\pi^{2i}}.
\eeq
Substituting this result into \Eq{g-M-G}, we find
\beq
\frac{g_M}{n^M}=(M!)^{2}\alpha^{M^2-1}\left(\sqrt{\dfrac{\tau}{2\gamma^{2}}}\right)^{M(M-1)}\prod\limits_{j=0}^{M-1}j!, \quad T_{d} \ll T\ll \gamma^{2}T_{d}.
 \label{correlation-T}
\eeq

\subsection{Discussions}
Table I shows the higher-order local correlations based on \Eq{correlation-C} and \Eq{correlation-T}.

 \begin{table}[h]
 \begin{center} 
 \begin{tabular}{|c|r|r|}
  \hline
   $\dfrac{g_{M}}{n^{M}}$& $T\ll T_{d}$ & $ T_{d}\ll T\ll \gamma^{2}T_{d}$\\
 \hline
  $\dfrac{g_{2}}{n^{2}}$ 
 &$\dfrac{4\alpha^{3}}{3}\left(\dfrac{\pi}{\gamma}\right)^{2}\left[1+\dfrac{1}{4}\left(\dfrac{\tau}{\pi}\right)^{2}\right]$ 
  & $\dfrac{2\tau}{\gamma^{2}}$ \\
 &&\\
  $\dfrac{g_{3}}{n^{3}}$ 
&$\dfrac{16\alpha^{8}}{15}\left(\dfrac{\pi}{\gamma}\right)^{6}\left[1+\dfrac{3}{2}\left(\dfrac{\tau}{\pi}\right)^{2}\right]$
  & $\dfrac{9\tau^{3}}{\gamma^{6}}$\\
  &&\\
  $\dfrac{g_{4}}{n^{4}}$ 
 &$\dfrac{1024\alpha^{15}}{2625}\left(\dfrac{\pi}{\gamma}\right)^{12}\left[1+5\left(\dfrac{\tau}{\pi}\right)^{2}\right]$
  & $\dfrac{108\tau^{6}}{\gamma^{12}}$\\
  &&\\
   $\dfrac{g_{5}}{n^{5}}$ 
 &$\dfrac{65536\alpha^{24}}{1157625}\left(\dfrac{\pi}{\gamma}\right)^{20}\left[1+\dfrac{25}{2}\left(\dfrac{\tau}{\pi}\right)^{2}\right]$
  & $\dfrac{4050\tau^{10}}{\gamma^{20}}$\\
  &&\\
   $\dfrac{g_{6}}{n^{6}}$ 
 &$\dfrac{16777216\alpha^{35}}{5615638875}\left(\dfrac{\pi}{\gamma}\right)^{30}\left[1+\dfrac{105}{4}\left(\dfrac{\tau}{\pi}\right)^{2}\right]$
  & $\dfrac{546750\tau^{15}}{\gamma^{30}}$\\
  &&\\
   $\dfrac{g_{7}}{n^{7}}$ 
&$\dfrac{4294967296\alpha^{48}}{79500599553375}\left(\dfrac{\pi}{\gamma}\right)^{42}\left[1+49\left(\dfrac{\tau}{\pi}\right)^{2}\right]$
  & $\dfrac{602791875\tau^{21}}{2\gamma^{42}}$\\
  &&\\
   $\dfrac{g_{8}}{n^{8}}$ 
  &$\dfrac{70368744177664\alpha^{63}}{219470547636040325625}\left(\dfrac{\pi}{\gamma}\right)^{56}\left[1+84\left(\dfrac{\tau}{\pi}\right)^{2}\right]$
  & $\dfrac{759517762500\tau^{28}}{\gamma^{56}}$\\
  &&\\
  \hline
  \end{tabular}
\caption{Higher-order correlations of bosons in two temperature regimes. The general forms of the correlation functions are given by Eqs. (\ref{correlation-C})
and (\ref{correlation-T}).}\label{table1}
\end{center}
 \end{table}
 

In Ref.~\cite{dmg_gvs}, Gangardt and Shlyapnikov used the leading term of the wave function to calculate the local correlations of 1D bosons.
They  used Jacobi polynomials and the moments of the distribution to evaluate the left hand side of \Eq{corre8} in Ref.~\cite{dmg_gvs}.
They calculated low-order correlation functions $g_2$ and $g_3$ in the temperature regime   $T\ll T_{d}$.
However, the coefficient of the temperature term in their $g_3$ disagrees with our result of $g_3$ in Table I. 
Here we have  considered the fact that  the $1/c$ correction to the wave function of the strongly interacting 1D bosons 
leads to a statistical correction to the pure Fermi statistics in their  calculation of the correlation function. 
 We were able to calculate any higher-order correlation functions in a wider range of temperaturs. 
With the help of the GES with $\alpha=1-2/\gamma$,  the higher-order local correlation functions which we obtained provide sub-leading order corrections (cp.\ Table I).
When $\alpha=1$,  correlations for the strongly interacting bosons,  which correspond to the Tonks-Girardeau  gas with
a pure Fermi statistics, were already calculated in Ref.~\cite{dmg_gvs} at zero temperature. 
From Table I  we have 
\begin{eqnarray}
 \frac{g_{2}}{n^{2}}&= &\dfrac{4\alpha^{3}}{3}\left(\dfrac{\pi}{\gamma}\right)^{2}\left[1+\dfrac{1}{4}\left(\dfrac{\tau}{\pi}\right)^{2}\right]\label{b2},\\
 \frac{g_{3}}{n^{3}}&=& \dfrac{16\alpha^{8}}{15}\left(\dfrac{\pi}{\gamma}\right)^{6}\left[1+\dfrac{3}{2}\left(\dfrac{\tau}{\pi}\right)^{2}\right]\label{b3}.
\end{eqnarray}
Besides the statistical  parameter $\alpha$  corrections, explicit formulas of the higher-order correlation functions are presented in Table I.
 
\begin{figure}[h]
\begin{tabular}{cc}
 \includegraphics[scale=0.95] {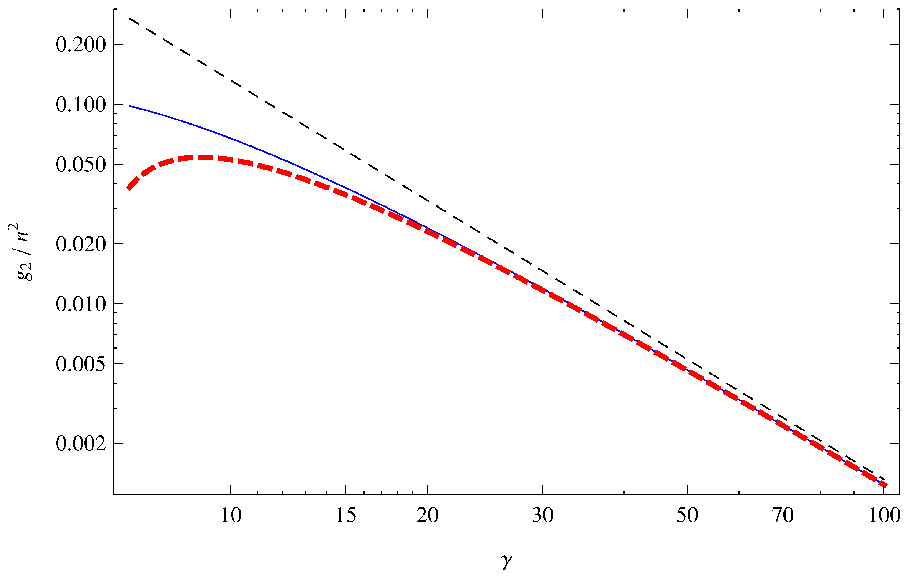}& \includegraphics[scale=0.95]{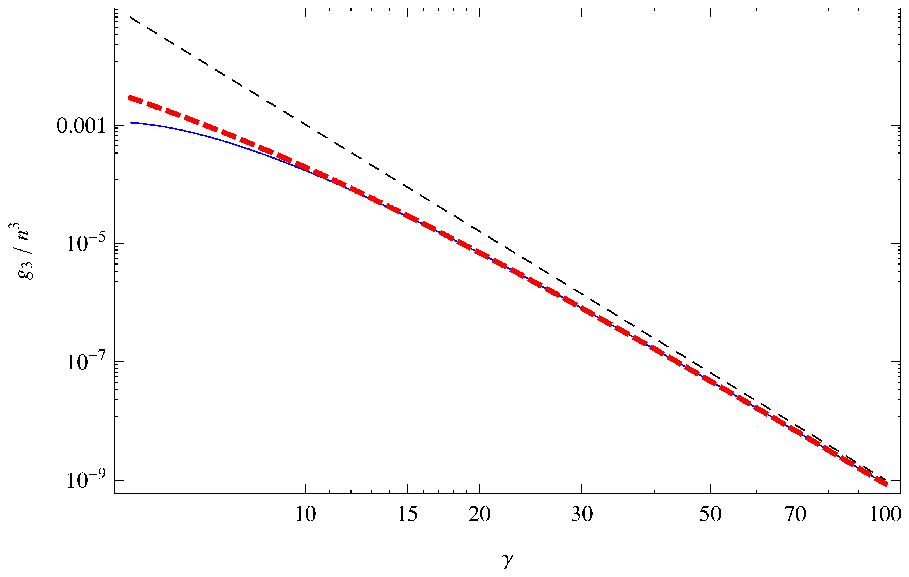}\\
 (a) & (b)
 \end{tabular}
 \caption{Double-logarithmic plots of the  $2$-body and $3$-body local correlation functions at zero temperature. These correlation functions  measure the probabilities  of finding two and three  particles at the same place, respectively. (a) A good  agreement between our result (\ref{b2}) (thin blue solid line)  and  the exact Bethe Ansatz result  \eqref{c2}  (thick red dashed line)  is observed. The  thin black  dashed line is the leading order of $g_2$ given in  \cite{dmg_gvs}.
  (b) An excellent  agreement between our  result (\ref{b2})  (thin blue solid line) and  the  approximate result   \eqref{vadi} (thick red dashed line)  is observed. The  thin black  dashed line is the leading order of $g_3$ given in  \cite{dmg_gvs}. }\label{fg1}
\end{figure}

 In addition, Wang {\em et.al.} \cite{wang} have analytically obtained the finite temperature local pair correlations for the strong coupling Bose gas at quantum criticality using the
polylog function in the framework of the TBA equations.  In the Luttinger liquid phase, their result \cite{wang} reduces to 
\begin{equation}\label{c2}
 \frac{g_{2}}{n^{2}}=\dfrac{4}{3}\left(\dfrac{\pi}{\gamma}\right)^{2}\left[1-\dfrac{6}{\gamma}+\dfrac{T^{2}}{4\pi^{2}(\hbar^{2}n^{2}/2m)^{2}}\right],
\end{equation}
which coincides with our result (\ref{b2}). 
FIG \eqref{fg1}(a) shows comparisons of the  $2$-body correlations represented in Eq.\ (34) in Ref.~\cite{dmg_gvs}  and the results of  Eqs. \eqref{b2} and \eqref{c2}.
Although it is clear that \eqref{b2} and \eqref{c2} agree with  each other very well, the Eq.~(34) in Ref.~\cite{dmg_gvs} has a deviation from  them
at strong but finite interaction strengths.  

Moreover,  Kormos {\em et. al.} \cite{Kormos:2009,Kormos:2010} developed a different method to compute the local correlation functions  using the  Sinh-Gordon model. 
They mapped the Lieb-Linger Bose gas onto  the Sinh-Gordon model with certain parameter limits. From the $M$-particle form factor of the local operator of the Sinh-Gordon model, they obtained 
the explicit form of $g_2$ and $g_3$ at $T=0$, namely
\begin{eqnarray}
\frac{g_2}{n^2} &=& \dfrac{4}{3}\left(\dfrac{\pi}{\gamma}\right)^{2}\left[1-\dfrac{6}{\gamma}+\frac{1}{\gamma^2}\left(24-\frac{8}{5}\pi^2 \right)\right]+O(\gamma^{-5}),\\
\frac{g_3}{n^3}&=&\dfrac{16}{15}\left(\dfrac{\pi}{\gamma}\right)^{6}\left( 1-\frac{16}{\gamma}\right)+O(\gamma^{-8}).
\end{eqnarray}
The former coincides with the result given in \cite{Guan:2011}, whereas the latter  is consistent with our result (\ref{b3}) with $T=0$.  To our best knowledge, our general formula 
(\ref{correlation-C}) derived here  coincides with the known results of $g_2$ and $g_3$  \cite{maxim,sykes,vladimir,holzmann,kv_dm_pd_gv,kozlowski,eckart,ovidiu,korepin,vadim,dmg_gvs,dmg_gvs1,oleksandr,Kormos:2010,Kormos:2011,Pozsgay:2011,Garcia:2014,Xu:2015,Guan:2011}.  However, there is  no explicit analytical expression of the local correlation functions $g_M$  for $M> 3$ for  the 1D Bose gas in the literature.
See also a recent study \cite{Pozsgay:2011}.
Cheianov  {\em et.al.} obtained two approximate formulas for $g_{3}$ at medium-to-strong couplings~\cite{vadim} 
\begin{equation}\label{vadi}
 \frac{g_3}{n^3}\Big|_{N=\infty}=\begin{cases}
              \dfrac{0.705-0.107\gamma+5.08\times 10^{-3}\gamma^{2}}{1+3.41\gamma+0.903\gamma^{2}+0.495\gamma^{3}}, \quad 1\leq\gamma\leq 7,\\
              \\
              \dfrac{16\pi^{6}}{15\gamma^{6}}\dfrac{9.43-5.40\gamma+\gamma^{2}}{89.32+10.19\gamma+\gamma^{2}}, \quad 7\leq \gamma\leq 30.
             \end{cases}
\end{equation}
FIG \eqref{fg1}(b) compares the  results among (\ref{b3}),  (\ref{vadi}) and the Eq.\ (35) in \cite{dmg_gvs}. It clearly shows that
when $\gamma \geq 7$ there is a very good agreement between  our result (\ref{b3}) and  the approximate expression of Cheianov {\em et.al.} (\ref{vadi}). 
All three results of $g_{3}$  reach the same asymptote in the limit $\gamma\rightarrow\infty$.

\section{IV $M$-body wave function at zero collision energy}

In this section, we consider the wave function of the 1D interacting bosons at zero collision energy
and zero total momentum.  If one boson has zero total momentum, its wave function is proportional to
\beq
\phi^{(1)}(x_1)\equiv1.
\eeq
When two bosons collide with zero total momentum and zero energy, their wave function is proportional to
\beq
\phi^{(2)}(x_1,x_2)\equiv|x_1-x_2|+A,
\eeq
where
\beq
A=\frac{2}{c}.
\eeq
$(-A)$ is the so-called 1D scattering length.

If $M$ bosons collide with zero total momentum and zero energy, and if we restrict our attention to the case in which their wave function grows no faster than
$\rho^{M(M-1)/2}$ at large $\rho$, where $\rho$ is the overall size of the system of $M$ bosons, then their wave function is uniquely determined
up to a multiplicative constant. This wave function is proportional to $\phi^{(M)}(x_1,x_2,\cdots,x_M)$. When $x_1\le x_2\le\cdots\le x_M$, for our convenience we define 
\begin{align}
\phi^{(M)}(x_1,x_2,\cdots,x_M)&\equiv\lim_{k\to0}\frac{1!\,2!\,\cdots(M-1)!}{\prod_{i<j}(\I k_j-\I k_i)}\sum_p(-1)^p\bigg[\prod_{i<j}\Big(1-A\frac{\mathrm{i}(k_{pi}-k_{pj})}{2}\Big)\bigg]\nn\\
&\quad\times\exp(\mathrm{i}k_{p1}x_1+\cdots+\mathrm{i}k_{pM}x_M),
\end{align}
where  the limit is understood as follows: hold the ratio $k_1:\cdots:k_M$ constant, and let $k_1$, \dots, $k_M$ shrink to zero simultaneously.
$p$ is one of the $M!$ permutations, and $(-1)^p$ is the signature of the permutation.
$(-1)^p=+1$ for even permutations and $(-1)^p=-1$ for odd permutations.
In addition, we define $\phi^{(M)}(x_1,x_2,\cdots,x_M)$
to be completely symmetric under the exchanges of its arguments.

The few-body asymptotic Bethe Ansatz wave functions defined here will help us to understand important correlation effects of  the many-body systems.

We have the following explicit formulas at $x_1\le x_2\le\cdots$:
\begin{eqnarray}
\phi^{(1)}(x_1)&=&1,\\
\phi^{(2)}(x_1,x_2)&=&(x_2-x_1)+A,\\
\phi^{(3)}(x_1,x_2,x_3)&=&(x_2-x_1)(x_3-x_1)(x_3-x_2)\nn\\
&& +A\big[(x_2-x_1)^2+4(x_2-x_1)(x_3-x_2)+(x_3-x_2)^2\big]+3A^2(x_3-x_1)+\frac{3}{2}A^3,\\
\phi^{(4)}(x_1,x_2,x_3,x_4)&=&(x_2-x_1)(x_3-x_1)(x_3-x_2)(x_4-x_1)(x_4-x_2)(x_4-x_3)\nn\\
&&+A \left(\delta _2 \delta _3^4+\delta _4 \delta _3^4+2 \delta _2^2 \delta _3^3+2 \delta _4^2 \delta _3^3+12 \delta _2 \delta _4 \delta _3^3+\delta _2^3 \delta _3^2+\delta _4^3
   \delta _3^2+15 \delta _2 \delta _4^2 \delta _3^2\right.\nn\\
   &&\left.+15 \delta _2^2 \delta _4 \delta _3^2+4 \delta _2 \delta _4^3 \delta _3+12 \delta _2^2 \delta _4^2 \delta _3+4 \delta _2^3
   \delta _4 \delta _3+\delta _2^2 \delta _4^3+\delta _2^3 \delta _4^2\right)\nn\\
&&+A^2 \left(\delta _3^4+10 \delta _2 \delta _3^3+10 \delta _4 \delta _3^3+12 \delta _2^2 \delta _3^2+12 \delta _4^2 \delta _3^2+48 \delta _2 \delta _4 \delta _3^2+3 \delta _2^3
   \delta _3\right. \nn\\
   &&\left. +3 \delta _4^3 \delta _3+33 \delta _2 \delta _4^2 \delta _3+33 \delta _2^2 \delta _4 \delta _3+3 \delta _2 \delta _4^3+9 \delta _2^2 \delta _4^2+3 \delta _2^3
   \delta _4\right)\nn\\
&&+\frac{1}{2} A^3 \left(3 \delta _2^3+42 \delta _3 \delta _2^2+39 \delta _4 \delta _2^2+66 \delta _3^2 \delta _2+39 \delta _4^2 \delta _2+144 \delta _3 \delta _4 \delta _2+16
   \delta _3^3\right. \nn\\
  &&\left. +3 \delta _4^3+42 \delta _3 \delta _4^2+66 \delta _3^2 \delta _4\right)\nn\\
&&+3 A^4 \left(3 \delta _2^2+13 \delta _3 \delta _2+11 \delta _4 \delta _2+7 \delta _3^2+3 \delta _4^2+13 \delta _3 \delta _4\right)\nn\\
&& +\frac{9}{2} A^5 \left(3 \delta _2+4 \delta _3+3 \delta _4\right)+\frac{9}{2}A^6,
\end{eqnarray}
etc, where $\delta_j\equiv x_j-x_{j-1}$.
In general, $\phi^{(M)}(x_1,\cdots,x_M)$ is a homogeneous polynomial of $x_1$, \dots, $x_M$ and $A$ of degree $M(M-1)/2$
at $x_1\le x_2\le\cdots\le x_M$.
One can show that
\begin{align}
\phi^{(M)}(x_1,\cdots,x_M)&=\Big|\prod_{i<j}(x_j-x_i)\Big|\nn\\
&\quad+O(A^1x^{M(M-1)/2-1})+O(A^2x^{M(M-1)/2-2})+\cdots+O(A^{M(M-1)/2-1}x^1)\nn\\
&\quad+1!\,2!\,\cdots(M-1)!\,M!\,(A/2)^{M(M-1)/2}.
\end{align} 
In particular, we observe 
\beq
\phi^{(M)}(x,\cdots,x)=1!\,2!\,\cdots(M-1)!\,M!\,(A/2)^{M(M-1)/2}.
\eeq

\section{V. $M$-body short-distance correlation of the ideal Fermi gas}

Consider a spin-polarized 1D  ideal Fermi gas with number density $n$ and temperature $T$. It has
momentum distribution
\beq
f(k)=\frac{1}{1+\exp\big(\frac{\hbar^2k^2/2m-\mu}{k_BT}\big)},
\eeq
where $m$ is the mass of each fermion, $k_B$ is Boltzmann's constant, $\mu$ is the chemical potential and as before 
$n=\int_{-\infty}^\infty\frac{dk}{2\pi}f(k)$.
Using Wick's theorem, we find that
\begin{align}
\langle\Psi^\dagger(x_1)\cdots\Psi^\dagger(x_M)\Psi(y_M)\cdots\Psi(y_1)\rangle&=\sum_P(-1)^P\int_{-\infty}^\infty\frac{dk_1}{2\pi}\cdots\frac{dk_M}{2\pi}f(k_1)\cdots f(k_M)\nn\\
&\quad\times e^{-(\I k_1x_1+\cdots +\I k_Mx_M)+(\I k_{1}y_{P1}+\cdots+\I k_{M}y_{PM})},
\end{align}
where $\Psi^\dagger(x)$ and $\Psi(x)$ are respectively the fermion creation and annihilation operators.
When the separations between these $2M$ coordinates are much smaller than both the average inter-particle spacing and the thermal de Broglie wave length, we find
\begin{align}
\langle\Psi^\dagger(x_1)\cdots\Psi^\dagger(x_M)\Psi(y_M)\cdots\Psi(y_1)\rangle&=\frac{I_M}{[1!\,2!\cdots(M-1)!]^2M!}\prod_{i<j}(x_j-x_i)(y_j-y_i)\nn\\
&\quad+\text{(higher-order terms in the separations)},\label{non-local-correlation-F}
\end{align}
where
\beq\label{IMdef}
I_M\equiv\int\frac{dk_1}{2\pi}\cdots\frac{dk_M}{2\pi}f(k_1)\cdots f(k_M)\Big[\prod_{i<j}(k_j-k_i)\Big]^2.
\eeq

At zero temperature we may use \Eq{corre8} and the related formulas in Sec. III to deduce
\beq\label{IMconjecture}
I_M=\frac{\big(\prod_{j=1}^{M-1}j!\big)^2\,M!}{\big[\prod_{j=1}^{M-1}(2j-1)!!\big]^2\,(2M-1)!!}\pi^{M(M-1)}n^{M^2}.
\eeq

Let $\hat{n}(x)=\Psi^\dagger(x)\Psi(x)$ be the local number density operator. We find that
\beq\label{Mbodynon-local1}
\langle \hat{n}(x_1)\cdots\hat{n}(x_M)\rangle=\frac{I_M}{[1!\,2!\cdots(M-1)!]^2M!}\prod_{i<j}(x_j-x_i)^2
\eeq
plus higher-order terms in the separations, at small but nonzero separations.

\section{VI. $M$-body short-distance non-local correlation of the strongly repulsive Bose gas}

In this section we concentrate on the non-local $M$-body correlation function of  the 1D strongly repulsive Bose gas, with $\gamma=\frac{c}{n}=\frac{2}{nA}\gg1$ and in the temperature regime $T\ll \gamma^2T_d$, where $T_d\equiv\hbar^2n^2/(2mk_B)$ is the quantum degeneracy temperature,
and $n$ is the number density. In such a regime, when $x_1,\cdots,x_M$ are sufficiently close to each other, such that their maximum separation
is comparable to or less than $A$, but the remaining $(N-M)$ particles are not that close to them, the $N$-body wave function is approximately factorized as
\beq\label{factorize}
\psi(x_1,\cdots,x_M,x_{M+1},\cdots,x_N)\approx \phi^{(M)}(x_1,\cdots,x_M)\Phi(\bar{x},x_{M+1},\cdots,x_N)
\eeq
plus higher-order corrections, where $\bar{x}=(x_1+\cdots+x_M)/M$.

Assuming that
\beq
\int dx_1\cdots dx_N|\psi(x_1,\cdots,x_N)|^2=1,
\eeq
we have
\begin{align}\label{Mbodyreduced1}
\langle\Psi^\dagger(x_1)\cdots\Psi^\dagger(x_M)\Psi(y_M)\cdots\Psi(y_1)\rangle&=\frac{N!}{(N-M)!}\int dx_{M+1}\cdots dx_N\nn\\
&\quad\quad\times\psi^*(x_1,\cdots,x_M,x_{M+1},\cdots,x_N)\nn\\
&\quad\quad\times\psi(y_1,\cdots,y_M,x_{M+1},\cdots,x_N),
\end{align}
where $\Psi^\dagger(x)$ and $\Psi(x)$ are respectively the boson creation and annihilation operators.
When the maximum separation of $x_1,\cdots,x_M,y_1,\cdots,y_M$ is comparable to or less than $A$, we may substitute \Eq{factorize} into \Eq{Mbodyreduced1}
to obtain
\beq\label{Mbodyreduced2}
\langle\Psi^\dagger(x_1)\cdots\Psi^\dagger(x_M)\psi(y_M)\cdots\Psi(y_1)\rangle\approx B_M\phi^{(M)*}(x_1,\cdots,x_M)\phi^{(M)}(y_1,\cdots,y_M),
\eeq
where
\beq
B_M=\frac{N!}{(N-M)!}\int dx_{M+1}\cdots dx_N\big|\Phi(\bar{x},x_{M+1},\cdots,x_N)\big|^2.
\eeq
We have ignored the tiny difference between $\bar{x}$ and $\bar{y}$ in \Eq{Mbodyreduced2}.
If the total momentum of the system is zero, $B_M$ is independent of $\bar{x}$.

A special case of \Eq{Mbodyreduced2} is
\beq\label{Mbodynon-local2}
\langle\hat{n}(x_1)\cdots\hat{n}(x_M)\rangle\approx B_M\big|\phi^{(M)}(x_1,\cdots,x_M)\big|^2,
\eeq
if the coordinates $x_1,\cdots,x_M$ do not coincide.
Here $\hat{n}(x)$ is the local number density operator of the bosons.

When the separations between $x_1,\cdots,x_M$ are much larger than $A$, but much smaller than both the average inter-particle spacing and
the thermal de Broglie wave length, we can use our knowledge of $\phi^{(M)}$ to deduce that
$$
\langle\hat{n}(x_1)\cdots\hat{n}(x_M)\rangle\approx B_M\prod_{i<j}(x_j-x_i)^2.
$$
Comparing the above formula with our result for the ideal Fermi gas [see \Eq{Mbodynon-local1}], we get
\beq
B_M=\frac{I_M}{[1!\,2!\cdots(M-1)!]^2M!}.
\eeq
Therefore, when the separations between $x_1,\cdots,x_M, y_1,\cdots, y_M$ are much smaller than both the average inter-particle spacing
and the thermal de Broglie wave length, we get
\beq\label{Mbodyreduced3}
\langle\Psi^\dagger(x_1)\cdots\Psi^\dagger(x_M)\Psi(y_M)\cdots\Psi(y_1)\rangle\approx \frac{I_M}{[1!\,2!\cdots(M-1)!]^2M!}
\phi^{(M)*}(x_1,\cdots,x_M)\phi^{(M)}(y_1,\cdots,y_M),
\eeq
where $\phi^{(M)}$ is defined in Sec.~IV.
At zero temperature, $I_M$ is given by \Eq{IMconjecture}. At nonzero temperatures $T\ll\gamma^2T_d$, one can use
\Eq{IMdef} to calculate $I_M$. When $T\gtrsim\gamma^2T_d$, \Eq{Mbodyreduced3} breaks down.
We emphasize that \Eq{Mbodyreduced3} is a \emph{key result} of this paper.

When the above $2M$ coordinates are all equal, we get
\beq
\langle\big[\Psi^\dagger(x)\big]^M\big[\Psi(x)\big]^M\rangle\approx M!\,I_M(A/2)^{M(M-1)}.
\eeq

At zero temperature, using \Eq{IMconjecture} we find
\beq
\frac{\langle\big[\Psi^\dagger(x)\big]^M\big[\Psi(x)\big]^M\rangle}{n^M}=\frac{\big[\prod_{j=1}^M(j!)\big]^2}{\big[\prod_{j=1}^{M-1}(2j-1)!!\big]^2(2M-1)!!}\Big(\frac{\pi}{\gamma}\Big)^{M(M-1)}
+o(\gamma^{-M(M-1)})
~~~\text{at }T=0, \label{non-local-correlation-B}
\eeq
which is consistent with \Eq{M-proof}.

 \section{VII. Conclusions}
 \label{section-dis}

 Higher-order quantum correlations  reveal the quantum many-body effects in ultracold atomic gases~\cite{yanli}.
 In light of  Gangardt and Shlyapnikov's method for  calculating  the higher-order correlation functions of 1D bosons  with an infinitely strong interaction,  we  have rigorously calculated the $M$-body correlation function. It turns out that the quasimomentum distribution correction $\alpha=1-2/\gamma$   to the free fermions leads to the sub-leading terms in the $M$-body correlation functions at a large interaction strength. We have calculated the higher-order local correlation functions in terms of  the statistical parameter $\alpha$  and obtained $g_M$ explicitly for arbitrary $M$ with sub-leading order terms.
These results not only  recover the expressions for $g_2$ and $g_3$ with the sub-leading terms given in the literature \cite{dmg_gvs,dmg_gvs1,Kheruntsyan,Cazalilla,Kormos:2009}  but   also provide explicit forms of $g_M$ with arbitrary $M$   at zero and  nonzero  temperatures. 
 To our best knowledge, there is not yet another such analytical expression of the local correlation functions $g_M$  (\ref{correlation-C}) for $M> 3$  in the literature for  the 1D Bose gas \cite{footnote2}.
Moreover, we have explicitly calculated the short-distance non-local $M$-body correlation functions of the 1D  free fermions and the 1D strongly interacting bosons in Eqs.~\eqref{non-local-correlation-F} and \eqref{Mbodyreduced3}.
Our results provide new insights into the many-body correlations in quantum systems of interacting bosons
and noninteracting fermions.

\section{APPENDIX I}

In the strong coupling limit, $\gamma\gg1$, the higher-order correlation function is given by \Eq{corre-f},
\begin{equation*}
   g_{M}=\frac{N!}{(N-M)!}\frac{\int|\psi(0,\cdots,0,x_{M+1},\cdots,x_{N})|^{2}dx_{M+1}\cdots dx_{N}}{\int|\psi(x_{1},\cdots,x_{N})|^{2}dx_{1}\cdots dx_{N}}.
\end{equation*}

\subsection{Calculating the numerator of the formula }

The Bethe Ansatz energy eigenfunction  in the domain $0\le x_{1}\leq x_{2}\leq \cdots\leq x_{N}\le L$ is
\begin{align}\label{eqa5}
\psi(x_{1},x_{2},\cdots,x_{N})&=\sum\limits_{p}(-1)^p\Big[\prod\limits_{1\leq i< j\leq N}\Big(1+\frac{\I k_{p{j}}-\I k_{p{i}}}{c}\Big) \Big]\exp\Big(\sum\limits_{j=1}^{N}\I k_{p{j}}x_{j}\Big)\nn\\
&=\Big[\prod\limits_{1\leq i< j\leq N}\Big(1+\frac{\partial_{x_{j}}-\partial_{x_{i}}}{c}\Big) \Big]\phi^{(0)}(x_{1},\cdots,x_{N})
\end{align}
where
\beq
\phi^{(0)}(x_{1},\cdots,x_{N})\equiv\sum\limits_{p}(-1)^p\exp\Big(\sum\limits_{j=1}^{N}\I k_{pj}x_{j}\Big),\quad \text{for all } x
\eeq
is completely antisymmetric under the exchange of its arguments. The function
\beq
\prod\limits_{1\leq i\leq M;M+1\leq j\leq N}\Big(1+\frac{\partial_{x_{j}}- \partial_{x_{i}}}{c}\Big)\prod\limits_{M+1\leq i< j\leq N}\Big(1+\frac{\partial_{x_{j}}-\partial_{x_{i}}}{c}\Big)\phi^{(0)}(x_{1},\cdots,x_{N})
\eeq
 is still antisymmetric under the exchange of any two coordinates $x_{i}$ and $x_{j}$ satisfying $1\leq i<j\leq M.$ Being smooth, such a function must vanish like $\delta^{M(M-1)/2}$ (or even faster) when the coordinates $x_{1},\cdots,x_{M}$ are of the order $\delta$ and $\delta$ goes to zero. Therefore, in the domain $0\leq x_{M+1}\leq x_{M+2}\leq\cdots \leq x_{N}\leq L$ we have

\begin{align}\label{eqa6}
\psi(0,\cdots,0,x_{M+1},\cdots,x_N)&=\prod_{1\le i<j\le M}\frac{\partial_{x_j}-\partial_{x_i}}{c}
\prod_{1\le i\le M; M+1\le j\le N}\Big(1+\frac{\partial_{x_j}-\partial_{x_i}}{c}\Big)\nn\\
&\mspace{10mu}\prod_{M+1\le i<j\le N}\Big(1+\frac{\partial_{x_j}-\partial_{x_i}}{c}\Big)\phi^{(0)}(x_1,\cdots,x_N)\Big|_{x_1=\cdots=x_M=0}.
\end{align}

At strong coupling, $c\to\infty$, we expand \Eq{eqa6}  as
\beq
\psi(0,\cdots,0,x_{M+1},\cdots,x_{N}) =\chi^{(0)}(x_{M+1},\cdots,x_{N}) +\chi^{(1)}(x_{M+1},\cdots,x_{N}) +O(c^{-M(M-1)/2-2}).
\eeq
In the domain $0\leq x_{M+1}\leq x_{M+2}\leq\cdots\leq x_{N}\leq L$ we have
\begin{align}
\chi^{(0)}(x_{M+1},\cdots,x_{N}) &=c^{-M(M-1)/2}\Big[\prod\limits_{1\leq i <j\leq M}\Big(\partial_{x_{j}}-\partial_{x_{i}}\Big)\Big]\phi^{(0)}(x_{1},\cdots,x_{N})\Big|_{x_{1}=\cdots=x_{M}=0},\label{eqa2}\\
\text{and}\quad
 \chi^{(1)}(x_{M+1},\cdots,x_{N}) 
&=c^{-M(M-1)/2-1}\Big[-(N-M)\sum\limits_{l=1}^{M}\partial_{x_{l}}+\sum\limits_{l=M+1}^{N}(2l-N-1)\partial_{x_{l}}\Big]\nn\\
&\quad\Big[\prod\limits_{1\leq i <j\leq M}\Big(\partial_{x_{j}}-\partial_{x_{i}}\Big)\Big]\phi^{(0)}(x_{1},\cdots,x_{N})\Big|_{x_{1}=\cdots=x_{M}=0}\label{eqa3}.
\end{align}

Let
\beq
 \phi^{(\Delta)}(\epsilon,x_{M+1},\cdots,x_{N}) \equiv\Big[\prod_{1\le i<j\le M}(\partial_{x_j}-\partial_{x_i})\Big]\phi^{(0)}(x_{1},\cdots,x_{N})\Big|_{x_{1}=\cdots=x_{M}=\epsilon}.
\eeq
Then in the domain $0\leq x_{M+1}\leq x_{M+2}\leq\cdots\leq x_{N}\leq L$  \Eq{eqa2} and \Eq{eqa3} become
\begin{align}
\chi^{(0)}(x_{M+1},\cdots,x_{N}) &=c^{-M(M-1)/2}\phi^{(\Delta)}(0,x_{M+1},\cdots,x_{N}) ,\\
\text{and}\quad
 \chi^{(1)}(x_{M+1},\cdots,x_{N}) &=c^{-M(M-1)/2-1}\Big[-(N-M)\frac{\partial \phi^{(\Delta)}(\epsilon,x_{M+1},\cdots,x_{N}) }{\partial\epsilon}\Big|_{\epsilon=0}\nn\\
&\quad +\sum\limits_{l=M+1}^{N}(2l-N-1)\frac{\partial \phi^{(\Delta)}(0,x_{M+1},\cdots,x_{N}) }{\partial x_{l}}\Big].
\end{align}

Therefore, in the domain $ 0\leq x_{M+1}\leq x_{M+2}\leq\cdots\leq x_{N}\leq L$ we have
\begin{align}
|\psi(0,\cdots,0,x_{M+1},\cdots,x_{N})|^{2}&=|\chi^{(0)}|^{2}+\chi^{(0)*}\chi^{(1)}+\chi^{(0)}\chi^{(1)*}+O(c^{-M(M-1)-2})\nn\\
&=c^{-M(M-1)}|\phi^{(\Delta)}(0,x_{M+1},\cdots,x_{N})|^{2}\nn\\
&\quad+c^{-M(M-1)-1}\Big[-(N-M)\frac{\partial|\phi^{(\Delta)}(\epsilon,x_{M+1},\cdots,x_{N})|^{2} }{\partial\epsilon}\Big|_{\epsilon=0}\nn\\
&\quad +\sum\limits_{l=M+1}^{N}(2l-N-1)\frac{\partial| \phi^{(\Delta)}(0,x_{M+1},\cdots,x_{N})|^{2} }{\partial x_{l}}\Big]+O(c^{-M(M-1)-2}).
\end{align}

Let
\begin{align}
\rho^{(\Delta)}(\epsilon)&\equiv \int_{0\leq x_{M+1}\leq\cdots\leq x_{N}\leq L}| \phi^{(\Delta)}(\epsilon,x_{M+1},\cdots,x_{N}) |^{2}dx_{M+1}\cdots dx_{N},\\
\rho^{(\Delta)}_{l}(x_{l})&\equiv \int_{0\leq x_{M+1}\leq\cdots\leq x_{N}\leq L}| \phi^{(\Delta)}(0,x_{M+1},\cdots,x_{N}) |^{2}dx_{M+1}\cdots dx_{l-1}dx_{l+1}\cdots dx_{N},
\end{align}
where $M+1\leq l\leq N.$
Then
\begin{align}\label{eqa4}
 &\int_{0\leq x_{M+1}\leq\cdots\leq x_{N}\leq L}|\psi(0,\cdots,0,x_{M+1},\cdots,x_{N})|^{2}dx_{M+1}\cdots dx_{N}\nn \\
&=
c^{-M(M-1)}\rho^{(\Delta)}(0)
 +c^{-M(M-1)-1}\bigg\{-(N-M)\frac{\partial\rho^{(\Delta)}(\epsilon)}{\partial\epsilon}\Big|_{\epsilon=0}
+\sum\limits_{l=M+1}^{N}(2l-N-1)\Big[\rho^{(\Delta)}_{l}(L)-\rho^{(\Delta)}_{l}(0)\Big]\bigg\}\nn\\
&\quad\quad+O(c^{-M(M-1)-2}).
\end{align}
At leading order in $1/c$, the quasimomenta $k_{j}\;(1\leq j\leq N)$ may be approximated as $(2\pi/L)\times$integers. This implies that
$\rho^{(\Delta)}(\epsilon)$ is independent of $\epsilon$ at leading order in $1/c,$ and
\beq
\frac{\partial \rho^{(\Delta)}(\epsilon)}{\partial \epsilon}\Big|_{\epsilon=0}=O(1/c).
\eeq

Consideration of the volumes of the domain of integration indicate that
\begin{align}
\rho^{(\Delta)}_{l}(0) =0, &\quad \text{if} \; M+2\leq l \leq N,\\
\rho^{(\Delta)}_{l}(L) =0,&\quad \text{if}\; M+1\leq l\leq N-1.
\end{align}

Because of the smoothness and the complete antisymmetry of $\phi^{(0)}(x_{1},\cdots,x_{N}),$ it is easy to see that when $x_{M+1}\rightarrow 0,$ the function $\phi^{(\Delta)}(0,x_{M+1},\cdots,x_{N})$ vanishes like $x^{M}_{M+1}\rightarrow0.$
Consequently
\beq
\rho^{(\Delta)}_{M+1}(0)=0.
\eeq

Finally,
\begin{align}
\rho^{(\Delta)}_{N}(L) &=\int_{0\leq x_{M+1}\leq\cdots\leq x_{N-1}\leq L}|\phi^{(\Delta)}(0,x_{M+1},\cdots,x_{N-1},L)|^{2}dx_{M+1}\cdots dx_{N-1}\nn\\
&=\int_{0\leq x_{M+1}\leq\cdots\leq x_{N-1}\leq L}|\phi^{(\Delta)}(0,x_{M+1},\cdots,x_{N-1},0)+O(1/c)|^{2}dx_{M+1}\cdots dx_{N-1}\nn\\
&=\int_{0\leq x_{M+1}\leq\cdots\leq x_{N-1}\leq L}|O(1/c)|^{2}dx_{M+1}\cdots dx_{N-1}\nn\\
&=O(1/c^{2}).
\end{align}

Combining the above findings, we simplify \Eq{eqa4} as
\begin{align}
&\int_{0\leq x_{M+1}\leq\cdots\leq x_{N}\leq L}|\psi(0,\cdots,0,x_{M+1},\cdots,x_{N})|^{2}dx_{M+1}\cdots dx_{N}\nn\\
& = c^{-M(M-1)}\rho^{(\Delta)}(0)+O(c^{-M(M-1)-2})\nn\\
&=c^{-M(M-1)} \int_{0\leq x_{M+1}\leq\cdots\leq x_{N}\leq L}|\phi^{(\Delta)}(0,x_{M+1},\cdots,x_{N})|^{2}dx_{M+1}\cdots dx_{N}+O(c^{-M(M-1)-2}).
\end{align}

Multiplying  the above equation by $(N-M)!$, we get
\begin{align}
&\int_{0}^{ L}|\psi(0,\cdots,0,x_{M+1},\cdots,x_{N})|^{2}dx_{M+1}\cdots dx_{N}\nn\\
&=c^{-M(M-1)} \int_{0}^{ L}|\phi^{(\Delta)}(0,x_{M+1},\cdots,x_{N})|^{2}dx_{M+1}\cdots dx_{N}+O(c^{-M(M-1)-2}).
\end{align}

\subsection{Calculating the denominator of the formula}

At strong coupling, $c\to\infty$, we expand \Eq{eqa5}  as
\beq
\psi(x_1,\cdots,x_N)=\psi^{(0)}(x_1,\cdots,x_N)+\psi^{(1)}(x_1,\cdots,x_N)+O(c^{-2}),
\eeq
where in the domain $0\le x_1\le x_2\le\cdots\le x_N\le L$ we have
\begin{align}
 \psi^{(0)}(x_1,\cdots,x_N)&=\sum_p(-1)^{p}\exp\Big(\sum_{j=1}^N\I k_{p{j}}x_j\Big),\\
\text{and}\quad
 \psi^{(1)}(x_1,\cdots,x_N)&=\sum_{1\le i<j\le N}\frac{(\partial_{x_j}-\partial_{x_i})}{c}\psi^{(0)}(x_1,\cdots,x_N)\nn\\
&=\frac{1}{c}\sum_{j=1}^N(2j-N-1)\partial_{x_j}\psi^{(0)}(x_1,\cdots,x_N).
\end{align}
Therefore
\begin{align}\label{eqa7}
\int_0^L|\psi(x_1,\cdots,x_N)|^2dx_1\cdots dx_N&=N!\int_{0\le x_1\le\cdots \le x_N\le L}\big|\psi^{(0)}+\psi^{(1)}\big|^2dx_1\cdots dx_N+O(c^{-2})\nn\\
&=N!\int_{0\le x_1\le\cdots \le x_N\le L}\Big[|\psi^{(0)}|^2+\Big(\psi^{(0)*}\psi^{(1)}+\psi^{(0)}\psi^{(1)*}\Big)\Big]dx_1\cdots dx_N+O(c^{-2})\nn\\
&=N!\int_{0\le x_1\le\cdots \le x_N\le L}|\psi^{(0)}|^2dx_1\cdots dx_N+\frac{2(N!)}{c}\sum_{j=1}^N(2j-N-1)\re b_j+O(c^{-2}),
\end{align}
where
\beq
b_j\equiv\int_{0\le x_1\le\cdots \le x_N\le L}\psi^{(0)*}\partial_{x_j}\psi^{(0)}dx_1\cdots dx_N.
\eeq
Since the total momentum $(k_1+\cdots +k_N)$ must be an integer times $(2\pi/L)$, it is easy to see that
when $0\le x_1\le\cdots\le x_N\le L$, we have
\begin{align}
\psi^{(0)}(L-x_N,L-x_{N-1},\cdots,L-x_1)&=\sum_p (-1)^p\exp\Big[\I\sum_{j=1}^N k_{p{j}}(L-x_{N-j+1})\Big]\nn\\
&=\sum_p (-1)^p\exp\big(-\I\sum_{j=1}^N k_{p{j}}x_{N-j+1}\big)\nn\\
&=(-1)^r\psi^{(0)*}(x_1,x_2,\cdots,x_N),
\end{align}
where $(-1)^r$ is the signature of the reversal permutation $\{1,\cdots,N\}\to\{N,\cdots,1\}$.
In particular, $(-1)^r=+1$ if $\mathrm{mod}(N,4)=0$ or 1, and $(-1)^r=-1$ if $\mathrm{mod}(N,4)=2$ or 3.

From the above equation one can show that
\beq
b_{N+1-j}=-b_j^*.
\eeq
Define
\beq
\rho_j(x)=\int_{0\le x_1\le\cdots\le x_{j-1}\le x\le x_{j+1}\le\cdots\le x_N\le L}|\psi^{(0)}(x_1,\cdots,x_{j-1},x,x_{j+1},\cdots,x_N)|^2dx_1\cdots dx_{j-1}dx_{j+1}\cdots dx_N.
\eeq
It is easy to see that
\beq
2\re b_j=\int_0^L\frac{\partial\rho_j(x)}{\partial x}dx=\rho_j(L)-\rho_j(0).
\eeq

From the definition of $\rho_j(x)$ we can easily see that
\begin{align}
\rho_j(0) &=0, \quad \text{if} \; j\ge2,\\
\text{and}\quad \rho_j(L)&=0,\quad \text{if}\;j\le N-1.
\end{align}

It is also easy to see that
\beq
\rho_1(0)=\rho_N(L).
\eeq
Thus
\beq
2\re b_j=0,~~\text{if }2\le j\le N-1,
\eeq
and, assuming that $N\ge2$, we have
\begin{align}
2\re b_1 &=-\rho_N(L),\\
2\re b_N&=+\rho_N(L).
\end{align}
So, \Eq{eqa7} is simplified as
\beq
\int_0^L|\psi(x_1,\cdots,x_N)|^2dx_1\cdots dx_N=\int_0^L|\psi^{(0)}|^2dx_1\cdots dx_N+\frac{N!\,2(N-1)}{c}\rho_N(L)+O(c^{-2}).
\eeq
Let
\beq
\rho_\text{arb}(x)\equiv\int_0^L|\psi^{(0)}(x_1,\cdots,x_{N-1},x)|^2dx_1\cdots dx_{N-1}.
\eeq
Strictly speaking, $\rho_\text{arb}(x)$ depends on $x$. But in the large $c$ limit if we approximate $\psi^{(0)}$ by $\psi$, then
$\rho_\text{arb}(x)$ is approximately proportional to the local number density, which is a constant at thermal equilibrium.
Thus
\beq
\rho_\text{arb}(x)=\big[1+O(1/c)\big]\frac{1}{L}\int_0^L\rho_\text{arb}(y)dy=\big[1+O(1/c)\big]\frac{1}{L}\int_0^L|\psi^{(0)}|^2dx_1\cdots dx_N,
~~\text{if }0\le x\le L.
\eeq
On the other hand, it is easy to see that

\beq
\rho_N(L)=\frac{1}{(N-1)!}\rho_\text{arb}(L).
\eeq
Thus, in the limit $c\to\infty$ we have
\beq
\int_0^L|\psi(x_1,\cdots,x_N)|^2dx_1\cdots dx_N=\Big[1+\frac{2N(N-1)}{cL}+O(c^{-2})\Big]\int_0^L|\psi^{(0)}|^2dx_1\cdots dx_N.
\eeq

\paragraph{Acknowledgments } 
The authors thank M. Kormos and M. Rigol   for helpful discussions. 
This  work has been supported by  the  NNSFC under grant numbers 11374331 and by the key NNSFC grant No. 11534014.
The author ST is supported by the U.S. National Science Foundation CAREER award Grant No. PHY-1352208.
This work also  has been partially supported   by CAS-TWAS President's Fellowship for International PhD students. 
The author RR was  funded by Chinese Academy  of  Science President's International Fellowship Initiative grant No.2015VMA011.

\end{document}